\documentstyle{eps}

\received{xxx 00, 2000}
\revised{xxx 00, 2000}
\accepted{Nov 11, 2000}
\volumenumber{00}
\publishyear{2001}
\frompage{0001}
\topage{00xx}

\shorttitle{Shibata \& Tanuma 2001, Earth, Planets and Space (in press), 'Plasmoid-Induced-Reconnection and Fractal Reconnection'}
\title{
Plasmoid-Induced-Reconnection and Fractal Reconnection
}
\author{Kazunari Shibata$^1$ and Syuniti Tanuma$^2$}
\affiliation{$^1$ Kwasan Observatory, Kyoto University, Yamashina, 
Kyoto 607-8471 \\
$^2$ Solar-Terrestrial Environment Laboratory, Nagoya University, Honohara 3-13, Toyokawa, Aichi 442-8507}

\ls{1.0}
\abstract{
As a key to undertanding the basic mechanism
for fast reconnection in solar flares,
{\it plasmoid-induced-reconnection} and 
{\it fractal reconnection}
are proposed and examined. 
We first briefly summarize recent solar observations
that give us hints on the role of plasmoid 
(flux rope) ejections
in flare energy release.
We then discuss the plasmoid-induced-reconnection model, 
which is an extention of the classical two-ribbon-flare
model which we refer to as the CSHKP model.
An essential ingredient of the new model is the 
formation and ejection of a plasmoid which play
an essential role in the storage of magnetic 
energy (by inhibiting 
reconnection) and the induction of a strong inflow into reconnection region.
Using a simple analytical model, we show
 that the plasmoid ejection and acceleration are closely
coupled with the reconnection process, leading to a 
{\it nonlinear instability} for the 
whole dynamics that determines the macroscopic 
reconnection rate uniquely.
Next we show that
the current sheet tends to have a {\it fractal
structure} via the following process path:
tearing $\Rightarrow$ sheet thinning
$\Rightarrow$ Sweet-Parker sheet
$\Rightarrow$ secondary tearing
$\Rightarrow$ further sheet thinning
$\Rightarrow$ ...
These processes occur repeatedly at smaller scales
 until a microscopic
plasma scale (either the ion Larmor radius or the ion inertial length)
is reached where anomalous resistivity or
collisionless reconnection can occur. 
The current sheet eventually 
has a fractal structure with many plasmoids 
(magnetic islands) of different sizes.
When these plasmoids are ejected out of the current
sheets, fast reconnection
occurs at various different scales 
in a highly time dependent manner. 
Finally, a scenario is presented for fast reconnection
in the solar corona on the basis of above
{\it plasmoid-induced-reconnection in a fractal current sheet}. 
}
\begin{document}

\maketitle
\vspace*{-143mm}
{\Large\bf Shibata \& Tanuma 2001}

{\Large\bf Earth, Planets and Space (in press)}
\vspace*{143mm}
%\begin{center}
%(November 11, 2000)
%\end{center}
%\copyrighttext{}

\vspace*{-20mm}
\section{Introduction}

Recent numerical simulations (e.g., Ugai 1986, 1992, 
Scholer 1989, Biskamp 1986, Yan, Lee, Priest 1992, 
Yokoyama and Shibata 1994, Magara and Shibata
 1999,  Tanuma et al. 1999, 2000) have revealed that 
if the resistivity is spatially uniform, 
fast, steady-state Petscheck-type reconnection does not  
occur but instead slow, Sweet-Parker-type  
reconnection occurs. This holds especially when a 
strong inflow is imposed at the external boundary, and
the only way so far found to achieve a steady-state Petschek
configuration is to have a {\it localized 
resistivity}. The so called anomalous resistivity 
satisfies this condition.  However, there are a number
of questions about it. 

1. In order to produce anomalous resistivity, 
the current sheet thickness
must be as small as the ion Larmor radius
\footnote
{Although the physics of anomalous resistivity has not yet been
fully understood, it is known that  anomalous resistivity
occurs due to plasma turbulence which is produced by the microscopic
plasma instability, such as the lower hybrid drift instability,
the electrostatic ion cyclotron instability, and the ion sound
instability (e.g., Treumann and Baumjohann 1997). 
In the case of the lower
hybrid drift instability, the threshold of the instability is
$ v_d > v_{ion,th}$ where $v_d = j/(n_e e) $ is the electron-ion
drift velocity, and $v_{ion,th} = (kT/m_i)^{1/2}$
is the ion thermal speed.  This equation becomes equivalent to
$d < r_{L,ion}$ if we consider the pressure balance 
$p=2nkT \simeq B^2/8\pi$ 
between inside and outside of the current sheet, where
$d$ is the thickness of the current sheet.}
$$  r_{L,ion} = {m_i v_{th} c \over eB} 
       = 100 \Bigl({B \over {\rm 10 G}} \Bigr)^{-1}
        \Bigl({T \over 10^6 {\rm K}} \Bigr)^{1/2}    \ \ {\rm  cm}   
                            \eqno(1)$$ 
or the ion inertial length
$$  l_{in,ion} = c/\omega_{p,i}
            = 300 \Bigl({n \over 10^{10} {\rm cm}^{-3}
                       } \Bigr)^{-1/2}   \ \ {\rm cm},   \eqno(2)$$
both of which are of order of 1 m  in 
the solar corona. Since the size of solar flares
is typically $10^4$ km, there is a large gap between
the flare size and the necessary microscopic scale
to produce anomalous resistivity. How can such 
an enormous gap between macroscopic  and 
microscopic scales be reconciled in real flares?

\copyrighttext{}

2. Even if the anomalous resistivity (or localized
resistivity) is realized, what determines the 
reconnection rate?

Based on recent observations of solar flares and
numerical simulations, 
we try to give possible answers to above questions.
We argue that the key physics needed to answer the above 
questions is the global coupling
between plasmoid (flux rope) ejection and 
reconnection process. Since this coupling is scale free,
it can occur on any scale, constituting a fractal
reconnection process, which couples the 
macro- and micro-scales.

\section{Solar Observations: Flares and Plasmoid Ejections}

Yohkoh has revealed numerous indications of magnetic reconnection
in solar flares,
such as cusps, arcades, loop top hard X-ray (HXR) sources, X-ray jets
and so on
(e.g., Tsuneta et al. 1992a, Hanaoka et al. 1994, Masuda et al. 1994,
Forbes and Acton 1996, Shibata 1999). 
Furthermore, 
as has been predicted by some pioneers
(Hirayama 1991, Moore and Roumeliotis 1992), 
the association
of plasmoid (flux rope) ejections
with flares is much more common than had been
thought (e.g., Shibata et al. 1995, Nitta 1996, 
Ohyama and Shibata 1997, 1998, 2000, Tsuneta 1997,
Akiyama and Hara 2000).
This has led us to advocate 
a unified model of flares shown in Figure 1
(Shibata  et al. 1995, Shibata 1996, 1997, 1998, 1999).
Recent observations with SOHO/LASCO 
have also revealed a lot of evidence of 
flux rope and disconnection events in coronal mass ejections (CMEs)
(e.g., Dere et al. 1999, Simnet et al. 1997), and 
Yohkoh has shown that giant arcades formed after prominence eruptions
or CMEs
are physically similar to flare arcades even though their total X-ray
intensity is much lower than that of normal flares
(e.g., Tsuneta et al. 1992b, Hanaoka et al. 1994).
Figure 2 shows several examples of plasmoid (flux rope) ejections on the Sun 
from the largest scale in CMEs ($\sim 10^{11}$ cm) 
to the smallest scale in compact
flares ($\sim 10^{9}$ cm). The velocity of these plasmoids range from
a few 10 km/s to 1000 km/s, and their maximum values are 
comparable to the inferred coronal Alfv\'en speed ($\sim 1000$ km/s).
These images show that the magnetic reconnection
and associated plasmoid ejection universally occur on 
widely different scales. 

One of the interesting findings by Yohkoh concerning 
X-ray plasmoid ejections is that, in impulsive flares,  
{\it a plasmoid starts to be ejected slowly,  
long before the impulsive phase,
and then is rapidly accelerated during the
impulsive phase} 
 (Ohyama and Shibata 1997, 1998,
Tsuneta 1997; Fig. 3).
Similar behavior  has also been observed
for LDE flares and CME events 
(e.g., Kahler et al. 1988, Hundhausen 1999).

Another interesting finding from Yohkoh on X-ray plasmoid ejection 
is that {\it there is a positive 
correlation between the 
plasmoid velocity ($V_{plasmoid} \sim 30 - 400$ km/s) and the 
apparent rise velocity of the flare loop ($V_{loop} \sim
 4 - 20$ km/s)}
(Shibata et al. 1995):
$$ V_{plasmoid} \simeq (8 - 20) \times V_{loop}.  
                     \eqno(3)$$
This relation (though still very preliminary) suggests
that the plasmoid velocity is related to the reconnection
inflow speed, or vice-versa.  This is because 
the apparent rise motion of the flare loop
is coupled to the reconnection process.
Consequently, magnetic flux 
conservation leads to
$$ V_{loop} \simeq (B_{inflow}/B_{loop}) V_{inflow}.  
                      \eqno(4)$$ 
Morimoto and Kurokawa (2000) found a correlation
between the erupting velocity of H-alpha filaments 
(i.e., a plasmoid) and the thermal energy density of post-eruption
X-ray arcades. This also suggests that there is a physical relation between 
plasmoid velocity and inflow speed (reconnection rate).

\section{Role of Plasmoid: 
Plasmoid-Induced-Reconnection Model}

On the basis of these observations, Shibata (1996, 1997) proposed
a {\it plasmoid-induced-reconnection model}, which is an extension of the 
classical CSHKP (Carmichel 1964, Sturrock 1966,
Hirayama 1974, Kopp and Pneuman 1976) model and similar in spirit to
the model of Anzer and Pneuman (1982).  
In this model, the plasmoid ejection plays
a key role in triggering fast reconnection in two different ways (Fig. 1).  

1) {\it A plasmoid (flux rope) can store energy by inhibiting reconnection.}
A large magnetic island (plasmoid or flux rope) 
inside the current sheet 
is a big obstacle for reconnection.
Hence if an external force compresses the current sheet, magnetic energy 
can be stored around the current sheet.  
Only after the plasmoid is ejected out of the current sheet, will the
anti-parallel field lines be able to touch and 
reconnect. If a larger plasmoid is ejected, a larger energy
release occurs. 

2) {\it A plasmoid ejection can induce 
a strong inflow into the reconnection site.}
If a plasmoid is suddenly ejected out of the current sheet
at the velocity  $V_{plasmoid}$,   
an inflow must develop toward the
X-point in order to compensate for the mass ejected by the plasmoid,
as has been shown in many numerical simulations
(e.g., Forbes 1990, Yokoyama and Shibata 1994, 2000,
Magara et al. 1997, Tanuma et al. 2000; 
see also Fig. 4).
The inflow speed can be estimated from the 
mass conservation law (assuming incompressibility, for simplicity);
$$V_{inflow} \sim V_{plasmoid} W_{plasmoid}/L_{inflow},
                     \eqno(5)$$
where $W_{plasmoid}$ is the typical width of the 
plasmoid, and $L_{inflow}  (\geq W_{plasmoid})$
is  the typical vertical length of the inflow region.
In deriving equation (5), it is assumed that the mass flux
into reconnection region ($\sim L_{inflow} V_{inflow}$)
is balanced by the mass flux carried by the plasmoid
motion ($\sim V_{plasmoid} W_{plasmoid}$).   
Since the reconnection rate is determined by the inflow speed,
the ultimate origin of fast reconnection in this model
is the fast ejection of the plasmoid.
If the plasmoid ejection (or outflow) 
is inhibited in some way, then fast reconnection 
ceases (Ugai 1982, Tanuma et al. 2000, Lin and Forbes 2000). 

This model naturally explains 
(1) the strong acceleration of plasmoids during the impulsive (rise) phase of
flares (see Fig. 3 and next section), 
(2) the positive correlation between plasmoid velocity and the 
apparent rise 
velocity of flare loops (eqs. 3 and 5),
(3) the total energy release rate of flares and plasmoid ejections
(Shibata 1997),
and 
(4) the time scale of the impulsive (rise) phase
 for both impulsive flares
($\sim L_{inflow}/V_{plasmoid} \sim $
$10^4$ km/100 km/s $\sim$ 100 sec), and
 for LDE flares ($\sim 10^5$ km/100 km/s 
$\sim 10^3$ sec).

It is interesting to note that similar impulsive 
reconnection associated with plasmoid ejection 
(current sheet ejection) has also been observed in
laboratory experiments (Ono et al. 2000).

\section{Nonlinear Instability caused by 
Strong Coupling between Plasmoid Ejection (Acceleration)
 and Reconnection}

In this section, we examine the physical mechanism of 
the plasmoid-induced-reconnection in more detail. 
We consider a situation where reconnection has just begun and 
a plasmoid, with a length  $L_p$ and a width  $W_p$, has just 
started to form.
The reconnection generates a jet (with  
the Alfv\'en speed $V_A$) which 
collides with the plasmoid and accelerates it.
Thus the plasmoid speed increases with time, which
induces a faster inflow into the 
reconnection point (i.e. the X-point), thereby  
leading to yet faster reconnection and an even larger energy release rate. 
This, in turn, accelerates the plasmoid again, eventually
leading to a kind of nonlinear instability for the plasmoid ejection and
the associated reconnection. 

Let us estimate the plasmoid velocity in this process, by 
assuming  that the plasmoid
is accelerated solely by the momentum of the reconnection jet.
(Note that we do not deny the possibility of acceleration of plasmoid
by other mechanism such as global magnetic pressure. The purpose
of this section is simply to show how the momentum of the 
reconnection jet can accelerate the plasmoid.)
We also assume that the plasmoid density $\rho_p$ and the 
ambient plasma density $\rho$ are constant with time, for simplicity.
In absence of any appropriate time-dependent theory in a rapidly
evolving configuration, we assume that the steady state mass
conservation $V_i L_i = V_p W_p$ (equation (5)) is valid 
and also that all the mass flux ($V_i L_i$) convected into
the reconnection region (with length of $L_i$) are accelerated
up to Alfven speed as in Sweet-Parker or Petschek model.

We first consider the case in which 
the mass added to the plasmoid by the reconnection jet
is much smaller than the total mass of the plasmoid
(i.e.,  the plasmoid speed $V_p$ is much smaller
than the Alfv\'en speed $V_A$).
Equating the momentum added by the reconnection jet with the change of
momentum of the plasmoid, we have
$$ \rho_p L_p W_p { d V_p \over dt} = \rho V_i L_i V_A           
                                = \rho V_p W_p V_A    \eqno(6)$$
where we use the mass conservation relation for the inflow and 
the outflow, $   V_p W_p = V_i L_i  $ (eq. 5). 
(See Appendix for detailed derivation of the equation (6).) 
Physically, this means that the inflow is induced by the outflow (plasmoid
ejection).
This  is the reason why this reconnection is called 
{\it plasmoid-induced-reconnection}.

The equation (6) is easily solved to yield the solution
$$ V_p = V_0 \exp(\omega t)          \eqno(7)    $$
where  $V_0$ is the initial velocity of the plasmoid, and
$$    \omega = {\rho V_A \over \rho_p L_p }.  \eqno(8)$$
Thus,  the plasmoid velocity increases exponentially with time,
and the \lq \lq growth time'' ($1/\omega$) 
is basically of order of Alfv\'en time.
The inflow speed becomes 
$$ V_i = {W_p \over L_i} V_p = 
  {W_p V_0 \exp(\omega t) \over 
  L_i(0) + {V_0 \over \omega} (\exp(\omega t) - 1)}   \eqno(9) $$
If $W_p$ is constant, the inflow speed increases exponentially with time
in the initial phase, but tends to be a constant ($\simeq \omega W_p$)
in the late phase.

As time goes on, the mass added to the plasmoid by the jet increases and
eventually becomes non-negligible compared with the initial mass
(i.e., the plasmoid speed becomes non-negligible compared
with the Alfv\'en speed).
In this case, we obtain the solution (see Appendix for derivation):
$$ V_p = {V_A \exp(\omega t)  \over \exp(\omega t) - 1 + V_A/V_0}. 
                    \eqno(10) $$
Hence the plasmoid speed is saturated at around
$ t = t_c \simeq {1 \over \omega} \ln (V_A/V_0)  $
and hereafter tends to the Alfv\'en speed $V_A$ as time goes on.
The inflow speed becomes
$$ V_i = {W_p V_p \over L_i}$$
$$     = W_p {V_A \exp(\omega t)/(\exp(\omega t) + a)
          \over (V_A/\omega)\ln [(\exp(\omega t)+ a)/(1+a)] + L_i(0) } 
        \eqno(11) $$
where
$ a = V_A/V_0 - 1.     $
If $W_p$ is constant in time, the inflow speed gradually decreases 
in proportion to $1/t$ after $t_c$.
\footnote{
This kind of evolution occurs when 1) {\it the current
sheet length is limited} (Tanuma et al. 2000), 
2) {\it magnetic field 
distribution is non-uniform around the current sheet}
(Magara et al. 1997).}
On the other hand, if $W_p$ increases with time in proportion to $t$
after $t_c$,
the inflow speed becomes constant, 
$$ V_i = \omega W_p(t=0) = 
               {\rho V_A \over \rho_p L_p} W_p(t=0)   \eqno(12)$$ 
In this case, the reconnection becomes
steady, and the shape of the reconnection jet and plasmoid
becomes self-similar in time and space (e.g., Nitta et al. 2000,
Yokoyama and Shibata 2000). 

A typical solution for $W_p$ = constant is shown in Figure 5, which
reminds us of the observed relation between plasmoid height vs. hard X-ray
intensity (Fig. 3; Ohyama and Shibata 1997) and explains also the
numerical simulation results (Fig. 4; Magara et al. 1997) very well. 
It is noted here that
the hard X-ray intensity is a measure of either the electric field at
the reconnection point ($E \propto V_i B$) or
the energy release rate
($\propto$ Poynting flux  $ \propto V_i B^2/(4\pi)$).

% We have seen that the plasmoid can be accelerated by
% the local reconnection even if there is no global acceleration
% of the plasmoid by the magnetic pressure. 
% The acceleration of the plasmoid is 
% strongly coupled with the reconnection dynamics, leading to the
% nonlinear instability. 
% The maximum velocity of the plasmoid
% is the Alfven speed, and 
% the acceleration time is of order of the Alfven time
% $\rho_p L_p/(\rho V_A)$.
% This nonlinear dynamics determines the maximum reconnection rate
% uniquely (if the resistivity increases in accordance with
% dynamics); 
% the maximum inflow speed is $W_p \rho V_A/(\rho_p L_p)$.
% This phase 
% corresponds to impulsive phase of solar flares.
% Actual dynamics would be nonsteady {\it  bursty reconnection}
% (Priest and Forbes 1992) since this macroscopic 
% reconnection rate is not necessarily coincident with
% microscopic reconnection rate.

\section{Fractal Reconnection}

As we discussed in section 1, 
we have a fundamental question:  how can we reach
the small dissipation scale necessary for anomalous 
resistivity or collisionless reconnection 
in solar flares?
Also, even if we can reach such a small scale, is it true that
there is only one diffusion region with a thickness
of 100 cm (and with a length of 10 m or 100 m) in a solar
flare as expected from
 Petschek's steady state theory?  

The idea that the reconnection process is inherently turbulent,
involving a spectrum of different scales, has been around for
some time (see Ichimaru 1975, for examples). However, 
here we argue that a plasma with large magnetic Reynolds
number (occurs as in the solar corona, the interstellar medium,
or the  intergalactic medium) inevitably leads to
a fractal current sheet with  many magnetic islands of 
different sizes connecting macroscopic and microscopic
scales (Tajima and Shibata 1997,
Shibata et al. 1997, 1998, Tanuma et al. 2000).

Let us first consider the Sweet-Parker current sheet with
a thickness of $\delta_{n}$
 and a length $\lambda_{n}$. 
This current sheet becomes unstable to secondary tearing if
$$t_n \leq \lambda_n / V_A,  \eqno(13)$$
where $t_n$ is the growth time of the tearing instability
at maximum rate ($\omega_{max} \propto $
${k_{max}}^{-2/5}
{t_{dif}}^{-3/5} {t_{A}}^{-2/5}$ and
$k_{max} \propto (t_{dif}/t_A)^{-1/4}$, where
$\omega_{max}$ and $k_{max}$ are the maximum growth rate
and corresponding wave number),
$$ t_n \simeq (t_{dif} t_A)^{1/2} \simeq
\Big({\delta_n^2 \over \eta}{\delta_n \over V_A}
         \Big)^{1/2},   \eqno(14)$$
and $\lambda_n/V_A$ is the time for the reconnection flow
 to carry the perturbation out of the current sheet.  
(As for the theory of the secondary tearing in the Sweet-Parker
sheet, see e.g.,
Sonnerup and Sakai (1981), Biskamp (1992).)
That is, if $t_n > \lambda_n/V_A$, the tearing instability
is stabilized by the effect of flow.   
Using eq (13) and (14), we find
$$ \delta_n^3 \leq \eta V_A \Big( {\lambda_n \over V_A}
                            \Big)^2,  $$
i.e., 
$$ \delta_n \leq  \eta^{1/3} V_A^{-1/3} \lambda_n^{2/3}.  \eqno(15)$$
If this inequality is satisfied, the secondary tearing occurs,
leading  to the current sheet thinning 
in the nonlinear stage of the
tearing instability.
At this stage, the current sheet thickness is
determined by the most unstable wavelength of the secondary
tearing instability, i.e., 
$$\lambda_{n+1} \simeq 6 \delta_n R_{m*,n}^{1/4}
              = 6 \eta^{-1/4} V_A^{1/4} \delta_n^{5/4}  
           \leq 6 \eta^{1/6} V_A^{-1/6} \lambda_n^{5/6},   
              \eqno(16) $$
where 
$ R_{m*,n} = \delta_n V_A/\eta. $
The current sheet becomes thinner and thinner, and when
the current sheet thickness becomes
$$ \delta_{n+1} \leq
         \eta^{1/3} V_A^{-1/3} \lambda_{n+1}^{2/3}, \eqno(17)$$
further secondary tearing occurs, and the same process
occurs again at a smaller scale (Fig. 6). 
It follows from eq (16) and (17) that
$$ \delta_n \leq \Big({\eta \over V_A }\Big)^{1/6} 6^{2/3} 
            \delta_{n-1}^{5/6},       \eqno(18) $$
or
$$ {\delta_n \over L} \leq A \Big( {\delta_{n-1} \over L} \Big)^{5/6}.  
               \eqno(19) $$
where
$$ A = 6^{2/3} R_m^{-1/6},      \eqno(20) $$
and
$$ R_m = {L V_A \over \eta}.      \eqno(21) $$
This 
fractal process continues until the current sheet thickness
reaches the microscopic scale such as the ion Larmor radius
or ion inertial length.
The equation (19) leads to 
$$ {\delta_n \over L} = A^{6(1-x)} \Big({\delta_0 \over L} \Big)^{x},
              \eqno(22a) $$
where
$$ x = (5/6)^n.    \eqno(22b)     $$
From this, we can estimate how many secondary tearings 
are  necessary for the initial
macroscopic current sheet to reach the microscopic scale.
Taking the typical solar coronal values,
$ \delta_0 = 10^8$ cm, $  L = 10^9$ cm, $V_A = 10^8$ cm/s, $\eta = 10^4$
cm$^2$/s for $T = 10^6$ K, we find $R_m = 10^{13}$ and
$$ A \simeq 0.02.    \eqno(23)  $$
Since $\delta_n$ must be smaller than the typical
microscopic scale, e.g., 
the ion Larmor radius ($\sim 100$ cm),
we have 
$$  \delta_n/L < r_{L,ion}/L,  \eqno(24) $$
or
$$   (0.02)^{6(1-(5/6)^n)} (0.1)^{(5/6)^n}  < 10^{-7}.  $$
The solution of this inequality (see Fig. 7) is
$$    n \geq 6.   \eqno(25)     $$
That is, in the solar corona, six secondary tearings are necessary to
reach microscopic current sheet. 

What is the time scale of this fractal tearing?
The time  scale for the n-th tearing is
$$ t_n \simeq \delta_n^{3/2} (\eta V_A)^{-1/2}
         = (\delta_n/\delta_0)^{3/2} t_0,  
                       \eqno(26)  $$
where
$$    t_0 = \delta_0^{3/2}/(\eta V_A)^{1/2}.   \eqno(27) $$
Since equation (22) leads to
$$     \delta_n/\delta_0 \simeq A_0^{6(1-(5/6)^n)}, 
                      \eqno(28)  $$
where
$  A_0 = 6^{2/3} R_{m*,0}^{-1/6},  $
and
$  R_{m*,0} = \delta_0 V_A/\eta,  $
we find
$$ t_n \simeq A_0^{9(1-(5/6)^n)} t_0.   \eqno(29) $$
Thus we obtain
$$ t_n/t_{n-1} = A_0^{(3/2)(5/6)^{n-1}} 
         \leq A_0^{3/2}    \eqno(30)   $$
for $n \geq 1$.   
It follows from this equation that
$$ t_n \leq A_0^{3/2} t_{n-1} 
       \leq A_0^{(3/2)n}t_0.   \eqno(31)  $$
Consequently, the total time from the 1st (secondary)
tearing ($t_1$) 
to the n-th (secondary) tearing ($t_n$) becomes
$$ t_{total} = t_1+t_2+...+t_n \leq 
t_0 A_0^{3/2}
{1- A_0^{3n/2} \over 1- A_0^{3/2}}
     \leq t_0 A_0^{3/2}.      \eqno(32) $$
For typical coronal conditions (described above), 
this time scale becomes 
$$ t_{total} \leq 6 \times 10^{-3} t_0,   \eqno(33)$$
which is much shorter than the  time scale of
the 0-th tearing ($t_0$). Although the 0-th tearing time
is long ($\sim 3 \times 10^4 - 10^6$ sec 
for initial current 
sheet with $\delta_0 \sim 10^7 - 10^8$ cm), 
 the  nonlinear
fractal tearing time  is quite short (less than
$10^2 - 3 \times 10^3$ sec), so that the microscopic
scale is easily reached within a short
time as a result of the fractal tearing.

It should be stressed that the role of the fractal 
tearing is
only to produce a very thin current sheet with a microscopic
scale of order of the ion Larmor radius or the ion inertial length.
The fractal tearing does
 not  explain the main energy release in flares.
The main energy release is explained by the 
fast reconnection process which occurs 
 after the ejection of the large
scale plasmoid as we discussed before. 

\section{Summary : A Scenario for Fast Reconnection}

Let us summarize our scenario
of fast reconnection in the solar corona, which is 
illustrated in Fig. 8 and Fig. 9 (the latter
is from a numerical simulation by 
Tanuma et al. 2000 and it nicely illustrates
 a part of our scenario).
Our scenario can also be applied to  other hot 
astrophysical plasmas (e.g., stellar corona, 
interstellar medium, galactic halo, 
galactic clusters, and so on) for which 
 magnetic Reynolds number
 and the ratio of its characteristic scale length 
 to the ion Larmor radius (or ion 
inertial length) are very large. 

Initially we assume  the current sheet
whose thickness is much larger than the microscopic
plasma scale.  Such a current sheet is easily created by the
interaction of emerging flux with an overlying coronal
field (e.g., Heyvaerts et al. 1977,
Shibata et al. 1992, Yokoyama and Shibata 1995),
the collision of a moving bipole with 
other magnetic structure
(e.g., Priest et al. 1994), 
the global resistive MHD instability in a shearing 
arcade 
(e.g., Mikic et al. 1988, Biskamp and Welter 1989,
Kusano et al. 1995, 
Choe and Lee 1996, Magara et al. 1997, Choe and Chen
2000), or other related  mechanisms (e.g.,
Forbes 1990, Chen, Shibata, Yokoyama 2000).

If the current sheet length becomes longer than
the critical wavelength for the tearing mode
instability, the instability starts.
As the instability developes, it enters a nonlinear
regime which  makes the initial current sheet thinner and 
thinner. 
The current sheet thinning stops when the sheet 
thickness becomes comparable to that of the 
Sweet-Parker sheet, and thereafter the sheet
length increases with time. 
If the sheet length becomes longer than a critical
wavelength (eq. 13),
secondary tearing occurs. 
Even if the sheet has not
yet reached the Sweet-Parker state, 
it can become
unstable to the secondary tearing if the sheet thickness
satisfies the same condition (eq. 13).
Then the same process occurs again at a smaller
scale, and the system evolves into one that  is fractally 
structured. In this way, a microscopicly 
small scale (such as ion Larmor radius or ion inertial
length) can be reached within a short time.

Once a small scale is achieved, fast reconnection
occurs because anomalous resistivity can now set in.
It is also possible that 
 fast collisionless reconnection occurs 
with a nondimensional
reconnection rate of the order of 0.1-0.01 
at this small scale
(see  recent full particle
 simulations by, e.g., Drake 2000, Hoshino et al.
2000, Tanaka 2000, Horiuchi and Sato 2000).
Hence small scale magnetic islands (plasmoids)
created by 
small scale tearing are ejected at the Alfv\'en speed and
collide with other islands to coalesce with each other,
thereby making bigger islands (plasmoids). 
 This coalescing process itself also occurs 
with a fractal nature (Tajima and Shibata 1997). 

It should be noted that the ejection (acceleration)
of plasmoids (flux rope with axial field in 3D space)
 can enhance the inflow into the
reconnection point, creating a positive feedback, i.e.,
nonlinear instability (as we outlined in section 4).
This determines the macroscopic reconnection rate
which may be smaller or larger than the microscopic
reconnection rate.  If the macroscopic reconnection
rate (inflow speed) is larger than the 
microscopic reconnection 
rate, the magnetic flux is accumulated
around the diffusion region, leading to intermittent
fast reconnection (Lee and Fu 1986, Kitabata et al. 1996,
Schumacher and Kliem 1996, 
Tanuma et al. 1999, 2000; Fig. 9). On ther other hand, if the 
macroscopic reconnection rate is smaller than the
microscopic reconnection rate, the reconnection
may continue in a quasi-steady state. 
However, 
there may be large amplitude perturbations 
around the reconnection point, so that 
it would be difficult to maintain quasi-steady
reconnection. The reconnection  
would be very time dependent with intermittent
reconnection and ejection of plasmoids with 
various sizes created by fractal reconnection.
Petschek's slow shocks are also formed in a very time dependent
manner (e.g., Yokoyama and Shibata 1994, Tanuma et al. 2000).
The local macroscopic 
reconnection rate can be much larger than
the average reconnection rate and is determined by
the macroscale dynamics, 
i.e., {\it plasmoid-induced-reconnection}.
In this case, the time dependence is essential for
determining the reconnection rate.

Since this process is scale free, we have fractal
structure in the global current sheet.
The greatest energy release occurs when the
largest plasmoid is ejected. 
This may correspond to the impulsive phase of
flares. 
The time variation of the reconnection rate 
(and the total energy release rate) associated with 
ejection of plasmoids with various sizes
 is also fractal. That is, 
the power spectrum of the time variation of the
reconnection rate and the energy release rate
show a power-law distribution.  
This may correspond to the 
fragmented light curves of solar X-ray 
and radio emissions in the impulsive phase of flares
(e.g., Benz and Aschwanden 1992).

Quantitative proof  of 
 the fractal nature of the current sheet (especially in 3D geometry) 
 remains as an important subject for future   
numerical simulations and laboratory experiments, 
both of which will  have to be able to handle
much larger magnetic Reynolds number than they currently do
(i.e. $R_m \simeq 10^3-10^4$) in order to solve this 
fundamental problem.

\section*{Acknowledgement}

The authors would like to thank K. Dere, 
J. Drake, M. Hoshino, T. Magara, T. Morimoto, M. Ohyama, 
Y. Ono, E. N. Parker, M. Scholer, M. Tanaka, 
T. Terasawa, M. Ugai, M. Yamada, and T. Yokoyama
for fruitful discussions. They also thank T. Forbes and
C. Z. Chen for their careful reading of our manuscript and
their many useful comments and suggestions which are
very useful to improve the paper.
Numerical computations were carried out on VPP300/16R and(or) VX/4R
at the Astronomical Data Analysis Center of
the National Astronomical Observatory, Japan,
which is an inter-university research institute of astronomy operated
by Ministry of Education, Science, Culture, and Sports.

\section*{Appendix: Derivation of Equations (6) and (10)}

As we wrote in the text, we assume that all the mass 
convected into the reconnection region 
($\rho V_i L_i$ per unit time per unit length in 2D space) 
are accelerated to 
Alfv\'en speed $V_A$. Since such accelerated mass (reconnection jet)
collides with the plasmoid, 
it can accelerate the plasmoid. Denoting $V_p =$ plasmoid speed,
$M_p=$ plasmoid mass, 
 $\Delta M_p =$ mass
convected by the reconnection jet during a short time
$\Delta t$, which is equal to increase in plasmoid mass during $\Delta t$,
we obtain the conservation of momentum as
$$ \Delta M_p V_A + M_p V_p = (M_p + \Delta M_p) (V_p + \Delta V_p). 
             \eqno(A1)$$
Here the left hand side is the total momentum before collision,
and the right hand side is the total momentum after collision.
 If we neglect the term $\Delta M_p$ in the right hand
side of equation (A1) 
(i.e., if we assume $V_p \ll V_A$), we have
$$ M_p \Delta V_p = \Delta M_p V_A.     \eqno(A2)$$
The plasmoid mass ($M_p$) and the mass added to the plasmoid 
($\Delta M_p$) by the jet for a short time
$\Delta t$ are written as 
$$M_p =  \rho_p L_p W_p,           \eqno(A3)$$
$$\Delta M_p = \rho V_i L_i \Delta t,        \eqno(A4)$$ 
both of which are per unit length. 
Using these formulae, the equation (A2) becomes
$ \rho_p L_p W_p  \Delta V_p = \rho V_i L_i V_A \Delta t, $
which is equivalent to
$$ \rho_p L_p W_p {d V_p \over dt} = \rho V_i L_i V_A.  \eqno(A5)$$

Since we assumed that the mass injection into reconnection region
is induced by the plasmoid motion, i.e., 
$V_i L_i = V_p W_p $ (equation (5)), 
 the right hand side of 
 equation (A5) becomes equal to $\rho V_p W_p V_A$, so that
we get equation (6):
$$ \rho_p L_p W_p {d V_p \over dt} = \rho V_i L_i V_A
           =  \rho V_p W_p V_A.     \eqno(6)$$

Note that in deriving above formulae, we did not assume conservation
of kinetic energy. This is because  
some part of the kinetic energy is dissipated to heat the plasmoid, 
leading to increase in gas pressure (internal energy) of the plasmoid.
Although such enhanced gas pressure may accelerate the plasmoid
further, we neglected the effect of gas pressure in above treatment
for simplicity, since it is not easy to estimate how much fraction
of internal energy is converted to the kinetic energy of a plasmoid.

When $V_p$ grows to the value that 
cannot be neglected compared with $V_A$, 
 we cannot  neglect the term $\Delta M_p V_p$ 
in the right hand side of equation (A1). 
In this case,  the momentum conservation equation becomes
$ M_p \Delta V_p = (V_A - V_p) \Delta M_p. $
Combining this equation with equations (A3) and (A4), 
we get
$$  {d V_p \over dt} = {\rho V_p \over \rho_p L_p}(V_A - V_p). \eqno(A6)$$
If $\rho, \rho_p, V_A,$ and $ L_p$ are constant in time, 
the solution of this equation becomes equation (10):
$$ V_p = {V_A \exp(\omega t)  \over \exp(\omega t) - 1 + V_A/V_0}. 
                    \eqno(10) $$
Here $V_0$ is the initial velocity of a plasmoid at $t = 0$,
and $\omega = (\rho/\rho_p)(V_A/L_p)$.

\clearpage

Figure Captions

\bigskip

Fig. 1  A unified model of flares: 
{\it plasmoid-induced-reconnection model}
(Shibata et al. 1995).  This is an 
extention of a classical model of eruptive
solar flares, called the CSHKP model.

\bigskip

Fig. 2  Various plasmoids (flux rope) with different
scales observed on the Sun. 
(a) Coronal mass ejection (CME), the largest-scale
plasmoid on the Sun ($\sim 10^{11}$ cm)
observed with SOHO/LASCO on Nov. 1-2, 1997 (Dere et al. 1999).
These are running-difference images. The velocity of the CME is
140 - 240 km/s. 
(b) Large-scale X-ray plasmoid associated with
an LDE (long duration event) flare on Feb. 21, 1992
($\sim 10^{10}$ cm) observed with Yohkoh/SXT
(Hudson 1994, Ohyama and Shibata 1998).
The plasmoid velocity is about 100 km/s.
(c) Small-scale X-ray plasmoid associated with an 
impulsive flare ($\sim 10^{9}$ cm) 
observed with Yohkoh/SXT on Oct. 5, 1992 (Ohyama and Shibata
1998).These are negative images.
The velocity of the plasmoid is 250 - 500 km/s.
1'' corresponds to 726 km.

\bigskip

Fig. 3   Temporal variations of the height of an X-ray
 plasmoid and the 
hard X-ray intensity of an impulsive solar flare
on 11 Nov. 1993 observed with Yohkoh SXT and 
HXT (Ohyama and Shibata 1997).

\bigskip

Fig. 4   Temporal variations of both 
the reconnection rate (electric field at the 
reconnection-point)
and the height of the plasmoid (magnetic island)
for a typical result of 2.5D MHD numerical simulation
of magnetic reconnection induced by plasmoid (flux rope) 
ejection (Magara, Shibata, Yokoyama  1997). 
Units of the height, time, and electric field are
$L$ (a half length between footpoints of a sheared arcade loop),
$t_0 = L/C_{s0}$ ($C_{s0}$ is the sound speed $\sim 0.4 V_A$), and 
$E_0 = C_{s0} B_0/c$, respectively.
In a typical solar coronal condition, $L \simeq 5000$ km,
$V_A \simeq$ 1000 km/s, 
$E_0 \simeq 2 \times 10^4$ V/m,
and $t_0 \simeq$ 20 sec.

\bigskip

Fig. 5  Temporal variations of the
plasmoid velocity ($V_{plasmoid}$),
its height, and inflow velocity ($V_{inflow}$),
in an analytical model (eqs. 10 and 11) for the case of
$V_A/V_0 = 100$.
Units of the velocity, height, and time are
$V_A, L_p$ and $L_p/V_A$, respectively.

\bigskip

Fig. 6  Schematic view of fractal reconnection.

\bigskip

Fig. 7  The current sheet thickness ($\delta_n/L$) 
in the $n$-th secondary tearing (see eq. 22a).

\bigskip

Fig. 8   A scenario for fast reconnection.
I: The initial current sheet. 
II: The current sheet thinning in the nonlinear
stage of the tearing 
instability or global
resistive MHD instability. The current sheet thinning
stops when the sheet evolves to the Sweet-Parker 
sheet.  
III: The secondary tearing in the Sweet-Parker sheet.  
The current sheet becomes fractal
because of further secondary tearing as shown in Fig. 6.
IV:  The magnetic islands coalesce with each other to form
bigger magnetic islands. The coalescence itself proceeds
with a fractal nature. In the phases III and IV, the
microscopic plasma scale (ion Larmor radius or
ion inertial length) is reached, so that 
fast reconnection becomes possible at small scales,
V:  The greatest energy release occurs when the largest
plasmoid (magnetic island or flux rope) is ejected.
The maximum inflow speed ($V_{in}$ = reconnection rate) 
is determined by the velocity of the plasmoid ($V_{p}$).
Hence this reconnection is called 
{\it plasmoid-induced-reconnection}.

\bigskip

Fig. 9  Numerical simulation of reconnection triggered
by an MHD fast mode shock (Tanuma et al. 2000), illustrating
a part of our proposed scenario for fast reconnection
(Fig. 8): a) passage of the MHD fast
shock, b) current sheet thinning (in the nonlinear stage
of the tearing instability), c) the Sweet-Parker reconnection,
d) secondary tearing, e) Petschek reconnection 
as a result of the onset of anomalous resistivity. 
Slow shocks inherent to Petschek reconnection are formed.

\end{document}